# Grain Surface Classification via Machine Learning Methods

Hüseyin Duysak [*1], Umut Özkaya [2] and Enes Yiğit [3]

[1*] Department of Electrical Electronics Engineering, Engineering Faculty, Karamanoğlu Mehmetbey University, Karaman, Turkey (ORCID: 0000-0002-2748-0660)
[2] Department of Electrical Electronics Engineering, Engineering Faculty, Konya Technical University, Konya, Turkey (ORCID 0000-0002-9244-0024)
[3] Department of Electrical Electronics Engineering, Engineering Faculty, Karamanoğlu Mehmetbey University, Karaman, Turkey (ORCID: 0000-0002-0960-5335)

**Abstract**

In this study, radar signals were analyzed to classify grain surface types by using machine learning methods. Radar backscatter signals were recorded using a vector network analyzer between 18-40 GHz. A total of 5681 measurements of A scan signals were collected. The proposed method framework consists of two parts. First Order Statistical features are obtained by applying Fast Fourier Transform (FFT), Discrete Cosine Transform (DCT), Discrete Wavelet Transform (DWT) on backscatter signals in the first part of the framework. Classification process of these features was carried out with Support Vector Machine (SVM). In the second part of the proposed framework, two dimensional matrices in complex form were obtained by applying Short Time Fourier Transform (STFT) on the signals. Gray-Level Co-Occurrence Matrix (GLCM) and Gray-Level Run-Length Matrix (GLRLM) were obtained and feature extraction process was completed. Classification process was carried out with DVM. 10-k cross validation was applied. The highest performance was achieved with STFT+GLCM+SVM.

## 1. Introduction

Silos are primary storage tools for grain. Grain is a so important food source because it contains various nutrients. So, the main task of silos is to protect and increase the storage life of grain[1]. For these reasons, determining the amount of grain is important in many situations such as seed planning, trades, and inventory tracking. That's why, different level measurement systems having own specific characteristics have been used[2]. The most important issue in silo level measurement is the conical grain stack structures inside the silo. These conical structures occur after the grain is unloaded or filled from the silo as seen from Figure 1. Single-point level determination systems such as ultrasonic, and laser, which provide a perfect solution for level measurement of liquids. Determination of the silo level with single-measurement point cannot produce a good solution due to the conical grain structures inside silo. Errors of these systems can be reduced by placing more sensor[3]. In addition, laser and ultrasonic systems are affected by dust, especially after grain filling and unloading. On the other hand, radar-based level measurement system is not affected by harsh environmental conditions and so it can be preferred in dusty environments such as silos. Besides, the whole surface of the grain can be illuminated electromagnetically by a radar system consisting of a wide beam width antenna. Thus, the level information of the whole grain surface can be obtained. However, it is very difficult to determine the grain level information from the complex reflection signal caused by the metallic walls of the silo and the surface structures of the grain. In recent years, in order to overcome these problems, combination of signal processing and machine learning algorithm has become popular[1], [4]. In this study, Support Vector Machine (SVM) algorithm is proposed to determine the grain surface type in silo by using radar backscattering signal dataset. The learning and test dataset were constituted using the radar-based experimental system. A total of 5681 measurements were performed for different grain surface conditions and different amounts of grain.

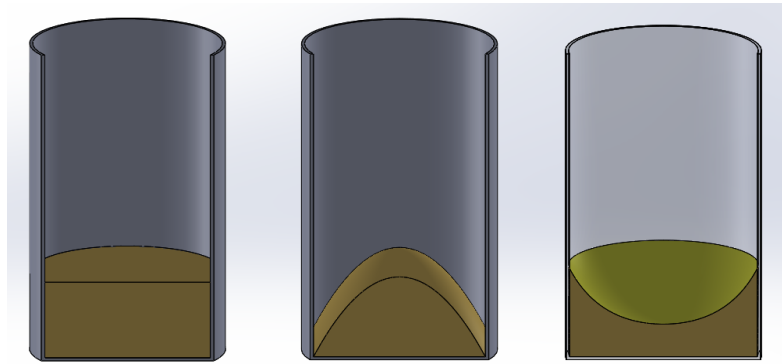

Figure. 1 Grain surface condition types in silo (a) levelled (b) peaked cone (c) inverted cone

The content of this paper is as follows: in the next chapter, the experimental measurement setup and proposed method are mentioned. In Chapter 3, and results are given. In the last section the study is summarized with conclusion remarks.

## 2. Material and Method

### 2.1. Experimental System

In order to perform silo level measurement, the experimental setup is constructed as shown in Figure 2. The experimental setup consists of model silo, vector network analyser (VNA), horn antennas, computer, grain and filling tube and volumetric cup. In this study, model silo is used instead of a commercial silo in order to carry out the experiments in the laboratory environment. Dimensions of model silo is calculated by scaling model method. In the scaling method, as the dimensions of the platform reduce, the measurement frequency increases. Dimensions of model silo corresponds to 8% of a commercial silo in [5]. Therefore, diameter of model silo is 36 cm. On the other hand, VNA is used to generate step frequency continuous wave radar (SFCWR) signal. SFCWR transmits radar signals in a certain frequency bandwidth and it collects the reflected signals from targets. Range resolution, $dz$ and the maximum unambiguous range of the radar, $R_{max}$ are calculated as follows,

$$dz = \frac{c}{2B} \quad (1)$$

$$R_{max} = N.dz \quad (2)$$

where c, $B$ and $N$ are speed of light, frequency bandwidth and the number of measurement frequencies, respectively.

The backscattering signal $S[k]$ reflected from the target at a distance R in SFCWR can be expressed as.

$$S[k] = \sum_{k=1}^{N} pe^{-jk2R} \quad (3)$$

where $k = \frac{2\pi f}{c}$ is wave number and $f = 1, 2, 3 ... N$ is frequency vector generated by SFCWR.

S [k] refers the backscattering signal of the whole surface of the grain in the frequency domain. The complex range profile S[r] is can be calculated by implementing the inverse Fourier Transform operation to S[k] as follows,

$$S[r] = \mathcal{F}^{-1}\{S[k]\} \quad (4)$$

On the other hand, two horn antennas were used to transmit and receive the SFCWR signal.

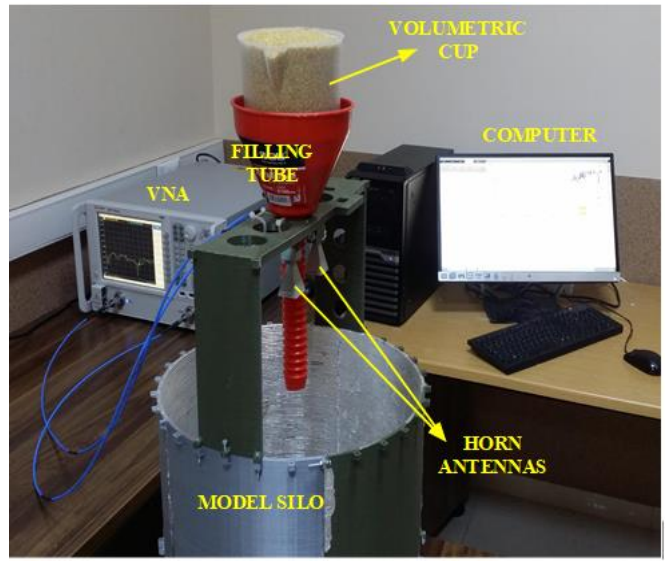

Figure. 2 Experimental System

### 2.2. Experiments and Construction of Dataset

Measurements were carried out in frequency range varying from 18 to 40 GHz with 301 frequency steps. Range resolution and maximum unambiguous range of the radar are calculated as 6.8mm and 2.04m via Eq.1 and Eq.2. On the other hand, according to scaling method, this frequency range corresponds to 1.5-3.2 GHz for commercial silo level measurement. As for experiments, different amounts of grain were filled and unloaded and so radar backscattering signal for three grain surface conditions (peaked cone, inverted cone, levelled) were obtained. In total, 5681 measurements were performed and recorded.

## 2.3. Proposed Framework

Two different parts in frameworks have been proposed for the classification of grain silo radar data. In the first part of proposed framework, different transformation methods are applied on the signals and different features are obtained. Within the scope of the second part in proposed framework, the responses of the signal data in the time and frequency domain are transformed by Short Time Fourier Transform (STFT) technique on the signals. By obtaining the size of the complex-sized STFT Gray level feature extraction methods were applied on the magnitude of STFT matrix. Proposed framework is given in Figure 3.

In the concept of first part of proposed framework, Fast Fourier Transform (FFT) performs operations on Discrete Fourier Transform (DFT) [6]. Performance of FFT is fast and efficient. Discrete cosine transform (DCT) is a technique used to separate the components of the signal in the frequency domain [7]. DCT technique is applied as a data compression algorithm and also is expressed as a finite number of data series in terms of the sum of the cosine functions. The basis of Discrete Wavelet Transform (DWT) method is the process of separating the main signal into sub-components by passing it through high and low pass filters. The low frequency signal obtained after the first level is subjected to wavelet transform again and divided into its own sub-components. It continues until the expected or desired signals are obtained in this way [8]. First Order Statistical (FOS) features are obtained and the classification process is performed after transformation processes, FOS features are based on mean, variance, curvature, kurtosis, entropy and energy.

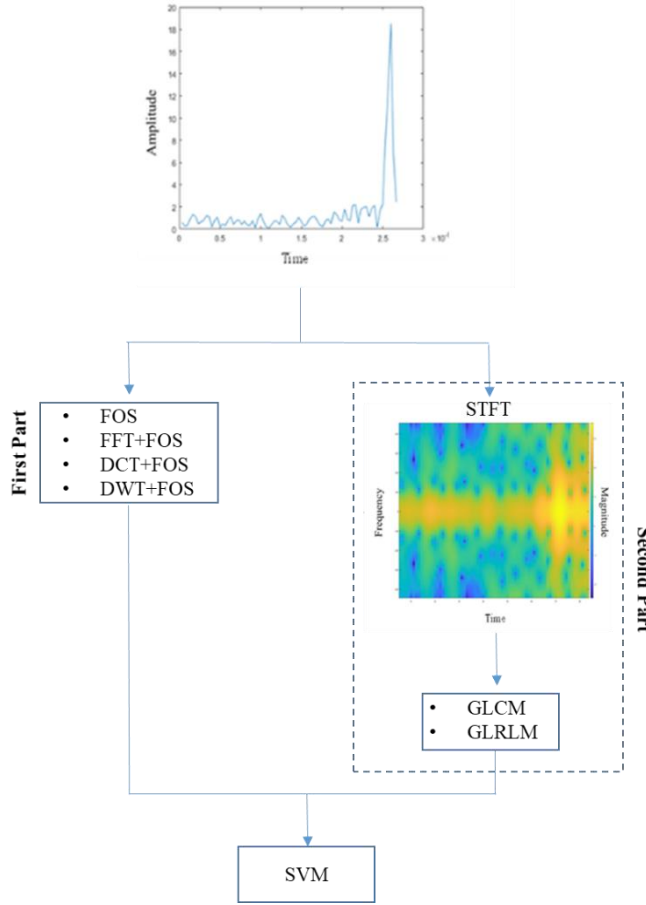

Figure. 3 Proposed Framework

Gray-Level Co-Occurrence Matrix (GLCM) and Gray-Level Run-Length Matrix (GLRLM) methods are used for feature extraction methods from magnitude of STFT matrix in the scope of second part for proposed method. In STFT, the non-stationary sign is divided into small pieces that can be considered as stationary in time. The sign is viewed through narrow windows and the sign inside the window is assumed to be stationary. Although the time-frequency representation of the signal is obtained, the width of the selected window plays an important role in the efficiency of the transformation. There is a resolution problem in STFT related to the window width [9].

In statistical texture analysis in two-dimensional data, computation is performed by taking into account the distribution density of the data. GLCM derives second order statistical features from two-dimensional data. Another alternative texture analysis method is GLRLM which evaluates similarities of each data in certain direction [10]. SVM was preferred for analysis and classification of features. SVM has gained popularity due to its ability to classify noisy and high dimensional data. SVM is a statistical classification method developed for classification and regression analysis. The background of SVM is basically to distinguish between the two classes of data in an optimal way. In classification process, the most suitable hyper plane line is found by using the training data. Then, classification is made according to which side of the test data is located on the boundary line. In order to find the most suitable hyperplane line, two plane lines forming the boundaries and parallel to the hyperplane are determined [11].

## 3. Results and Discussion

Classification process has been implemented by using two different concepts within the scope of the proposed framework. Six different classification metrics have been selected to evaluate the performance of classification processes using machine learning methods. These are sensitivity (SEN), specificity (SPE), accuracy (ACC), precision (PRE), F1-score and Matthews Correlation Coefficient (MCC).

$$Accuracy = (TP+TN)/(TP+FN+TN+FP) \tag{5}$$

$$Sensitivity = TP/(TP+FN) \tag{6}$$

$$Specificity = TN/(TN+FP) \tag{7}$$

$$Precision = TP/(TP+FP) \tag{8}$$

$$F1-Score = (2\times TP)/(2\times TP+FN+FP) \tag{9}$$

$$MCC = \frac{TP\times TN - FP\times FN}{\sqrt{(TP+FP)(TP+FN)(TN+FP)(TN+FN)}} \tag{10}$$

Results for first and second part of proposed framework are given in Table 1 and Table 2. According to the results for the first part of the proposed framework, the highest performance was achieved with DWT + FOS + SVM method. The metric performance of this method is 96.96 ± 0.77% SEN, 98.43 ± 0.39% SPE, 97.29 ± 0.68% ACC, 97.68 ± 0.66% PRE, 97.29 ± 0.69% F1-Score and 95.86 ± 1.04% MCC. The lowest performance was shown by the FOS + SVM method with 74.64 ± 1.02% SEN, 87.64 ± 0.52% SPE, 78.69 ± 0.74% ACC, 80.12 ± 0.85% PRE, 76.03 ± 0.92% F1-Score and 65.80 ± 1.16% MCC. STFT + GLCM + SVM method achieved the highest performance within the scope of the proposed framework. The performance of this method is 97.47 ± 0.56% SEN, 98.67 ± 0.28% SPE, 97.48 ± 0.46% ACC, 97.42 ± 0.39% PRE, 97.44 ± 0.45% F1-Score and 96.11 ± 0.70% MCC.

*Table 1. Performance for First Part of Proposed Framework*

| Methods | Evaluation Metrics (%) | | | | | |
|---|---|---|---|---|---|---|
| | SEN | SPE | ACC | PRE | F1-Score | MCC |
| *FOS+SVM* | 74.64±1.02 | 87.64±0.52 | 78.69±0.74 | 80.12±0.85 | 76.03±0.92 | 65.80±1.16 |
| *FFT+FOS+SVM* | 80.01±1.51 | 90.42±0.71 | 82.82±1.32 | 82.75±1.72 | 80.88±1.55 | 72.28±2.23 |
| *DCT+FOS+SVM* | 81.43±1.15 | 91.07±0.55 | 83.98±1.05 | 84.07±1.34 | 82.33±1.14 | 74.26±1.71 |
| *DWT+FOS+SVM* | **96.96±0.77** | **98.43±0.39** | **97.29±0.68** | **97.68±0.66** | **97.29±0.69** | **95.86±1.04** |

*Table 2. Performance for Second Part of Proposed Framework*

| Methods | Evaluation Metrics (%) | | | | | |
|---|---|---|---|---|---|---|
| | SEN | SPE | ACC | PRE | F1-Score | MCC |
| *STFT+GLCM+SVM* | **97.47±0.56** | **98.67±0.28** | **97.48±0.46** | **97.42±0.39** | **97.44±0.45** | **96.11±0.70** |
| *STFT+GLRLM+SVM* | 70.57±1.52 | 86.01±0.82 | 75.63±1.43 | 75.48±1.87 | 70.99±1.68 | 59.92±2.49 |

## 4. Conclusions and Recommendations

In this study, a system for classification of radar-based grain surface is presented. Grain surface classification can be performed quickly and efficiently with the proposed method. The method proposed as a machine learning-based system can accurately classify the 3-D grain surface. In the literature, studies have been intensively focused on detection of grain amount than grain surface classification. Therefore, this problem remained unsolvable. This study showed that grain surface classification can be implemented with high performance. STFT + GLCM + SVM machine learning method showed the highest performance in accordance with obtained results. In future studies, performance can be increased by using deep learning methods.

## 4. Acknowledge

This work is supported by Karamanoglu Mehmetbey University, Projects of Scientific Investigation unit (grant numbers BAP 19-M-17 and BAP 22-M-18).